\documentclass[superscriptaddress,amsmath, nofootinbib]{revtex4}
\usepackage{amsfonts}
\usepackage{graphicx}
\usepackage[latin1]{inputenc}
\usepackage{amsmath}
\usepackage{amssymb}
\usepackage{euscript}
\usepackage{hyperref}
\begin{document}
%\draft

%\preprint{APS/123-QED}

\title{An alternative interpretation of the Grioli gyroscope suspension points.}
%%%% 

\author{Alexei A. Deriglazov }
\email{alexei.deriglazov@ufjf.br} \affiliation{Depto. de Matem\'atica, ICE, Universidade Federal de Juiz de Fora,
MG, Brazil}

\date{\today}% It is always \today, today,

\begin{abstract}
We present a detailed analysis of all possible regular precessions of a heavy asymmetric body with a fixed point not coinciding with the center of mass. The calculations are done in terms of the rotation matrix, by writing the  Euler-Poisson equations with all involved vectors parameterized in the Laboratory frame. It is shown that a regular precession is possible if the suspension point is chosen on the straight lines (lying in the principal plane) which are frontiers of the regions where, as the distance from the center of mass increases, the interchange of the intermediate and largest moments of inertia occurs. Like the spin of an electron in quantum mechanics, the frequency of regular precession in classical mechanics turns out to be rigidly fixed by two values, i.e., quantized.  
\end{abstract}

\maketitle %\noindent
%%%%%%{\bf DOI:}
%%%%%%%{\bf PACS numbers:} 11.10.Ef, 03.65.Ca \\
%%%%%%%%{\bf Keywords: Thomas Precession, Relativistic Spin, Noncommutative Geometry}

%\tableofcontents

%\newpage

\section{Introduction.}\label{1RP}   

The most general motion of an axially symmetric free ideal body is a regular precession, when the body rotates uniformly around an axis fixed in the body, which in turn uniformly precesses around another axis fixed in space \cite{Lei_1965, Landau_8, Yeh_22, AAD_RB}. This motion is fundamental in various branches, including gravity and space science (dynamics of asteroids, air vehicles and spacecrafts), and modern engineering (high-precision gyroscope measurements and control) \cite{G_1, G_2, G_3, G_4, G_5, G_6, G_7, G_8, G_9, G_10, G_11, G_12, G_13, G_14, G_15, G_16, G_17, G_18, G_19, G_20, G_21, G_22}. However, a real body cannot be exactly axially symmetrical, but is to a greater or lesser extent asymmetrical. Numerous studies have been devoted to the analysis of questions about the existence of regular precessions of the asymmetric body in different circumstances, their stability, and control \cite{Yeh_22}.

It is known that a free asymmetrical body can not experience a regular precession, and it would be difficult to expect that the situation could improve in the gravitational field. Moreover, analyzing this problem, Routh showed that regular precession with the gravity vector chosen as the precession vector is impossible \cite{Routh_1897}. However, reexamining this issue, Grioli reported a remarkable result about the possibility of a regular precession with an inclined precession axis \cite{Grioli_1947}. Gulyaev confirmed and clarified the Grioli's analysis \cite{Gul_1955}, and since then, a large number of studies have been devoted to the analysis of precessions in various situations, see the works \cite{Yeh_22,  Yeh_2017, Mark_2018, Gorr_2023, Olsh_2023} and references therein.  

From his analysis, Grioli concluded that regular precession is possible if the point of suspension of the body is chosen 
on a straight line passing through the center of mass and perpendicular to one of two circular sections of the inertia ellipsoid calculated at the center of mass. An asymmetric body with such a suspension point is called a Grioli gyroscope. 

This task is traditionally solved in one or another body-fixed frame, and then the results should be translated in the Laboratory (fixed in space) frame,  where the body is observed. We proceed in a more direct way, in terms of the rotation matrix, by writing and analysing the Euler-Poisson equations for all involved vectors parameterized in the Laboratory frame\footnote{It should be noted that in the rigid body dynamics there is a number of specific properties which are not always taken into account in the traditional formalism  when formulating the rigid-body laws of motion and applying them\cite{AAD_RB, AAD_23, AAD23_2, AAD23_3}. This led to the need to reconsider some classical problems, including the problem of the motion of a Lagrange top, and a dancing spinning top \cite{AAD_2023_13, AAD23_8, AAD23_5, AAD_2023_9}.}. In the present work, we use the formalism of rotation matrix \cite{AAD_RB, AAD_23} to obtain and describe all possible regular precessions of an asymmetric heavy body (there are only two of them). We then describe an interpretation of the suspension points of the Grioli gyroscope that is different from the one discovered by Grioli.
The final results of our analysis can be described as follows.  

An asymmetrical heavy body with a fixed point (asymmetric gyroscope, for short) generally can not experience regular precession.  In order to have this ability, it must be "prepared`` in a very special way (see Item I below). Then it must be placed in a gravitational field (taking into account the axial symmetry of the problem) in a unique way (see Item II below). 

{\bf I.}  We take the principal moments of inertia, calculated at the center of mass $G$, to be ordered as follows: 
\begin{eqnarray}\label{int.1.1}
A_g<B_g<C_g. 
\end{eqnarray}
The suspension point $O$ must be chosen in the plane of the smallest and largest axes of inertia, and lie on any one of two straight lines passing through the center of mass at the angle 
\begin{eqnarray}\label{int.2.1}
\cos \varphi_g=\pm\sqrt{\frac{B_g-A_g}{C_g-A_g}},
\end{eqnarray}
to the smallest axis. They are perpendicular to circular sections of the inertia ellipsoid.  Our alternative interpretation of these straight lines will be described in Sect. \ref{6RP}. They turn out to be frontiers of the regions where, as the distance from the center of mass increases, the interchange of the intermediate and largest moments of inertia occurs. 
\begin{figure}[t] \centering
\includegraphics[width=07cm]{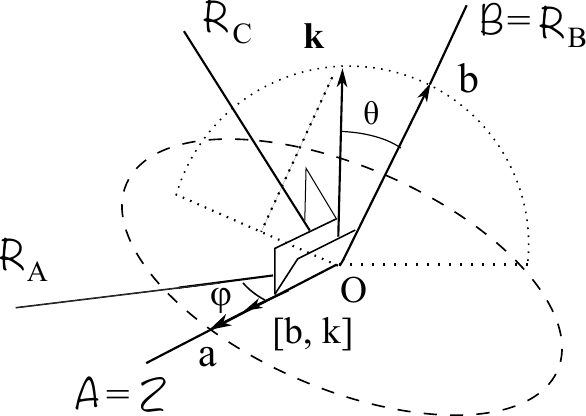}
\caption{The unique initial position of an asymmetric gyroscope that can experience a regular precession. The axes ${\EuScript R}_B$ and ${\bf k}$ lie in the plane of the paper sheet. ${\bf a}$ is the rotation vector. ${\bf b}$ is the precession vector.}\label{Reg_Prec_1}
\end{figure}

{\bf II.} Let us consider the gravity field with the acceleration of gravity equal $g>0$ and directed opposite to the constant 
unit vector ${\bf k}$, see Figure \ref{Reg_Prec_1}.
By $f$ we denote the number $f\equiv gL\mu$, where $L$ is the distance from the center of mass to the fixed point, and $\mu$ is the total mass of the body.  We take the principal moments of inertia, calculated at the suspension point $O$, to be ordered as follows: 
\begin{eqnarray}\label{int.1}
A<B<C. 
\end{eqnarray}
The corresponding inertia axes are denoted by ${\EuScript R}_A$, ${\EuScript R}_B$, and ${\EuScript R}_C$. 
% ETO PRAVILNO!!!!!
As we show below, Eq. (\ref{int.2.1}) implies that the angle between the dynamical smallest axis of inertia and the center-of-mass axis ${\EuScript Z}=OG$ in this case is 
\begin{eqnarray}\label{int.2}
\cos \varphi=\pm\sqrt{\frac{B-A}{C-A}}.
\end{eqnarray}
At the initial instant, we will first place our body in the gravitational field with the intermediate axis of inertia ${\EuScript R}_B$ along the vector ${\bf k}$. Now let us tilt the body in the plane of the paper sheet so that the gravity vector will have the angle\footnote{We point out that $A+C-B=2g_2$, where $g_2$ is an element of the mass matrix, see page 40 in \cite{AAD_RB}.}
\begin{eqnarray}\label{int.3}
\cos \theta=\frac{A+C-B}{\sqrt{(C-B)(B-A)+(A+C-B)^2}},
\end{eqnarray}
with the intermediate axis. The unit vector along this axis, which has an acute angle 
with ${\bf k}$, we denote as ${\bf b}$. Further, let us denote\footnote{See the end of this section for our notation.} by ${\bf a}$ the unit vector along $[{\bf b}, {\bf k}]$. Finally, we rotate the body about the 
vector ${\bf b}$ until the center-of-mass axis lies along ${\bf a}$. If in the configuration obtained the projection of  
vector ${\bf k}$ and the axis ${\EuScript Z}$ are in different quadrants of the plane of inertia axes ${\EuScript R}_A$ and ${\EuScript R}_C$, our gyroscope is ready for the regular precession, see Figure \ref{Reg_Prec_1}. If not, the body should be rotated 180 degrees about the axis ${\EuScript R}_A$.

{\bf Theorem 1.} The asymmetric gyroscope described in Items I and II will experience a regular precession with the same frequency $\alpha$ of rotation and precession. In this case, the body rotates about the center-of-mass axis, which in turn precesses 
about the axis ${\bf b}$ (as we saw above, the latter represents a copy of the intermediate axis taken at the initial instant and then fixed in space). The rotation about ${\bf a}$ and precession about ${\bf b}$ are simultaneously either counter-clockwise or clockwise. The frequency is fixed in a unique way by the dynamical principal moments and the number $f$ as follows:
\begin{eqnarray}\label{int.4}
\alpha^2=\frac{f}{\sqrt{(C-B)(B-A)+(A+C-B)^2}}.
\end{eqnarray}
The two described cases (that is, the counter-clockwise or clockwise movements) exhaust all possible regular precessions of an asymmetric gyroscope. The rotation matrix of these two regular precessions is the product of two pure rotations (\ref{1.1}): $R_{\pm}(t)=R^{\bf b}(\pm\alpha t)R^{\bf a}(\pm\alpha t)$, with ${\bf a}$, ${\bf b}$, and $\alpha$ as described above.

Let us enumerate some affirmations implied by Theorem 1. 

\noindent 1. For the case of an asymmetric gyroscope, its center of mass cannot precess about the gravity vector. 

\noindent 2. During a regular precession, the intermediate inertia axis at some instant will necessarily coincide with the precession vector ${\bf b}$. In our theorem, this instant was chosen as the initial one. At this instant, the position of the body (that is, the directions of inertia and center of mass axes) in the gravitational field is fixed in a unique way as described in Item II. 

\noindent 3. The only possible regular precession is the orthogonal regular precession: $({\bf a}, {\bf b})=0$. 

\noindent 4. Regular precession is only possible with coinciding frequencies of rotation and precession. 

\noindent 5. Total mass $\mu$ and the distance $L$ to the center of mass affect only the frequency (\ref{int.4}). The direction of the unit vector to the center of mass, lying on the axis of rotation, does not affect precession: for a gyroscope with an initial position in which ${\bf z}(0)=+{\bf a}$ 
or ${\bf z}(0)=-{\bf a}$, the rotation matrix turns out to be the same. 

In the remainder of this article, we confirm the necessary and sufficient conditions formulated in Items I and II, and then prove Theorem 1.

The work is organized as follows. In Sect. \ref{2RP} we present the necessary notation and describe the notion of a regular precession in terms of the rotation matrix. In Sect. \ref{3RP}, for the sake of completeness, we prove that an asymmetrical free body can not experience a regular precession. In Sect. \ref{4RP} we start our analysis of an asymmetric gyroscope and obtain a number of necessary conditions of a regular precession using the Laboratory frame, adapted with the rotation and precession axes. In Sect. \ref{5RP} we finish this analysis by using the Laboratory frame adapted with the inertia axes.  In Sect. \ref{6RP} we analyze the dependence of the principal moments and axes of inertia on the choice of the suspension point and, on this basis, present an alternative interpretation of Grioli's suspension points. Like the spin of an electron in quantum mechanics, the frequency of regular precession in classical mechanics turns out to be rigidly fixed, i.e., quantized. In the concluding section \ref{7RP}, we discuss this curious analogy.

%Like the spin of an electron, the quantization of which is embedded in the postulates of quantum mechanics,
%The uniqueness of regular precession

\begin{center}
{\bf Notation.} 
\end{center}

We mainly use the formalism and notation from \cite{AAD_RB, AAD_23}. We emphasize that our calculations are carried out with all quantities parameterized in the Laboratory frame, where the body is observed. 

Latin letters $i, j, k, \ldots$ are used to label coordinates. Repeated Latin indices are summed unless otherwise indicated: $\epsilon_{ijk}a^j b^k=\sum_j \sum_k \epsilon_{ijk}a^j b^k$. 

Vectors are denoted using bold letters. Given a vector, the corresponding axis (unoriented straight line along the vector) will be denoted by cursive letters: ${\bf a}\rightarrow {\EuScript A}$,  ${\bf R}_1(t) \rightarrow{\EuScript R}_1(t)$, and so on. 

A dot over any quantity means the time derivative of that quantity: $\dot{\bf a}=\frac{d{\bf a}}{dt}$.

Notation for the scalar product:  $({\bf a}, {\bf b})=a_i b_i$. Notation for the vector product: $[{\bf a}, {\bf b}]_i=\epsilon_{ijk}a_j b_k$, 
where $\epsilon_{ijk}$ is Levi-Civita symbol in three dimensions, with $\epsilon_{123}=+1$. 

We use the notation $A^T$ for a transposed matrix (vector).

\noindent ${\bf e}_1=(1, 0, 0)^T$,  ${\bf e}_2=(0, 1, 0)^T$, ${\bf e}_3=(0, 0, 1)^T$ - orthonormal basis vectors of the Laboratory system. \par

\noindent $R(t)$ - $3\times 3$ rotation matrix with the elements $R_{ij}(t)$.  Its columns ${\bf R}_i(t)=(R_{1i}, R_{2i}, R_{3i})^T$ form basis vectors of the body-fixed frame. Its lines ${\bf G}_j(t)=(R_{j1}, R_{j2}, R_{j3})$ represent coordinates of the Laboratory vectors ${\bf e}_j$ in the body-fixed frame. \par

\noindent  ${\bf a}$ - rotation vector\par

\noindent ${\bf b}$ - precession vector. \par 

\noindent ${\bf z}(0)$ - unit vector from fixed point to the center of mass at $t=0$.  \par

\noindent ${\boldsymbol\omega}$ - vector of angular velocity of the body. \par

\noindent $\omega _k=-\frac 12\epsilon_{kij}(\dot RR^T)_{ij}$ - components of the vector of angular velocity in Laboratory frame. \par

\noindent $\Omega_k=-\frac 12\epsilon_{kij}(R^T\dot R)_{ij}$ - components of the vector of angular velocity in body-fixed frame. \par

\noindent $I$ (or $J$) - tensor of inertia with the elements $I_{ij}$ (or $J_{ij}$). \par

\noindent $A_g, B_g, C_g$ - principal moments of inertia calculated at the center of mass point $G$. Asymmetrical body: $A_g<B_g<C_g$. Symmetrical body: $A_g=B_g < C_g$. Totally symmetrical body: $A_g= B_g= C_g$.\par

\noindent $A<B<C$ - principal moments of inertia calculated at the fixed point $O$. The corresponding inertia axes 
are ${\EuScript R}_A$, ${\EuScript R}_B$, ${\EuScript R}_C$, while unit vectors of the body-fixed frame along them are ${\bf R}_A$,  ${\bf R}_B$, ${\bf R}_C$. The plane of the smallest and largest inertia axes is called the principal plane. \par 

\noindent When the ordering of inertia moments is not important or not known, they are denoted by $I_1$, $I_2$, $I_3$, the corresponding axes are ${\EuScript R}_1$, ${\EuScript R}_2$, ${\EuScript R}_3$, while unit vectors of the body-fixed frame along them are ${\bf R}_1$,  ${\bf R}_2$, ${\bf R}_3$. \par

\noindent Dynamically asymmetrical body: $I_1\ne I_2\ne I_3$. Dynamically symmetrical body: $I_1= I_2\ne I_3$.

\section{Regular precession in terms of rotation matrix.}\label{2RP}

Consider a rigid body moving in space. According to Euler's rotation theorem, the movement of any of its points ${\bf y}(t)$  can be presented as follows: 
\begin{eqnarray}\label{1.3}
{\bf y}(t)={\bf y}_0(t)+ R(t){\bf x}(0).
\end{eqnarray}
In this expression, the term ${\bf y}_0(t)$  describes the motion of the center of mass, $R(t)$ is an orthogonal matrix called the rotation matrix, and ${\bf x}(0)$ are coordinates of the point ${\bf y}(t)$ relative to the center of mass at $t = 0$. Euler's theorem thereby reduces the problem of the motion of a free body to finding the temporal evolution of the rotation matrix. The latter contains all information on the evolution of the body in the Laboratory (fixed in space) frame, where the body is observed, and therefore is the main quantity that is required to be found.  As it should be, the rotation matrix satisfies second-order differential equations that can be derived by following the standard prescriptions of classical mechanics \cite{AAD_RB, AAD_23}. 
Using the Hamiltonian formalism, these equations can be rewritten as a first-order system by introducing auxiliary variables (conjugate momenta). They are known as the Euler-Poisson equations, see Eqs. (\ref{6.7}), (\ref{6.8}) below. 
In the case of a rigid body, the auxiliary variables turn out to be  
\begin{eqnarray}\label{1.1.1}
\Omega _k=-\frac 12\epsilon_{kij}(R^T\dot R)_{ij}. 
\end{eqnarray}
They have a geometric interpretation, being components of instantaneous angular velocity ${\boldsymbol\omega}(t)$, 
written in the body-fixed frame ${\bf R}_i(t)=R(t){\bf e}_i$. They are related as follows: ${\boldsymbol\omega}(t)=R(t){\boldsymbol\Omega}(t)$. Then ${\boldsymbol\omega}(0)={\boldsymbol\Omega}(0)$. 

If two rotation matrices $R_2(t)$ and $R_1(t)$ differ only in the shift of the time variable: $R_2(t+\delta)=R_1(t)$, $\delta=const$, then they describe the same movement, but with the initial data chosen at different moments in time.

Let ${\bf a}=(a_1, a_2, a_3)^T$ be unit vector with the components $a_i$ in a Laboratory system with the basis ${\bf e}_1, {\bf e}_2, {\bf e}_3$. Let us draw ${\bf a}$ with the initial point at the origin of the Laboratory. Rotation of a point ${\bf x}$ about ${\bf a}$ with angular frequency $\alpha$ is described by an orthogonal matrix as follows: ${\bf x}(t)=R^{\bf a}(\alpha t){\bf x}(0)$. Its explicit form is  
\begin{eqnarray}\label{1.1}
R^{\bf a}_{ij}(\alpha t)=\delta_{ij}\cos\alpha t+(1-\cos\alpha t)a_i a_j-\epsilon_{ijk}a_k\sin\alpha t. 
\end{eqnarray}
Let us agree to use only positive frequencies, $\alpha>0$. This implies counter-clockwise rotation around ${\bf a}$ when viewed from the end of the rotation vector. Clockwise rotations are taken into account by using of  $\alpha>0$ with $-{\bf a}$ instead of ${\bf a}$.

Using the definition (\ref{1.1.1}), the components of angular velocity corresponding to the pure rotation (\ref{1.1}) turn out to be constants 
\begin{eqnarray}\label{1.2}
{\boldsymbol\Omega}^{\bf a}=\alpha {\bf a}. 
\end{eqnarray}
Besides this, from ${\boldsymbol\omega}^{\bf a}(t)=R^{\bf a}(\alpha t){\boldsymbol\Omega}^{\bf a}$ we get ${\boldsymbol\omega}^{\bf a}={\boldsymbol\Omega}^{\bf a}$. 

The movement of a rigid body is called regular precession if its rotation matrix is the product of pure rotations with non-collinear vectors ${\bf a}$ and ${\bf b}$ and constant frequencies $\alpha$ and $\beta$ 
\begin{eqnarray}\label{1.4}
R(t)=R^{\bf b}(\beta t) R^{\bf a}(\alpha t). 
\end{eqnarray}
For the body's points which at $t=0$ lie along the axis ${\bf a}$ (if any), we get ${\bf x}(t)=R^{\bf b}(\beta t) R^{\bf a}(\alpha t){\bf x}(0)=R^{\bf b}(\beta t){\bf x}(0)$. That is, they experience pure rotation around ${\bf b}$. The movement of other points is a combination of two rotations. At the instant $t$ it can be thought as an instantaneous rotation about the axis $R(t){\bf a}=R^{\bf b}(\beta t){\bf a}$ fixed in the body, which in turn precesses about the axis ${\bf b}$ fixed in space.  In accordance to this, the vector ${\bf a}$  (and sometimes $R(t){\bf a}$) is called the rotation vector\footnote{It is known also as the figure vector \cite{Yeh_22}.}, while the space-fixed vector ${\bf b}$ is the vector of precession. 

According to Eq. (\ref{1.2}), for pure rotations the angular velocity vector is constant, both in space and in the body. Angular velocity of the regular precession is no longer a constant vector. Computing the components $\Omega_i(t)$ according to the equation (\ref{1.1.1}) for the matrix (\ref{1.4}), we get
\begin{eqnarray}\label{1.5}
{\boldsymbol\Omega}(t)={\boldsymbol\Omega}^{\bf a}+R^{{\bf a} T}(\alpha t){\boldsymbol\Omega}^{\bf b}=\alpha{\bf a}+\beta R^{{\bf a} T}(\alpha t){\bf b}.
\end{eqnarray}
Note however, that the following quantities turn out to be time-independent: 
\begin{eqnarray}%\label{1.6}
{\boldsymbol\Omega}^2(t)={const}=(\alpha+\beta)^2, \label{1.6}  \\
({\bf a}, {\boldsymbol\Omega}(t))=const=\alpha+\beta({\bf a}, {\bf b}). \label{1.7}
\end{eqnarray}
Consider ${\boldsymbol\Omega}(t)$ as the position vector of a three-dimensional space. Then the relations (\ref{1.5})-(\ref{1.7}) 
mean that the point ${\boldsymbol\Omega}(t)$ moves in the plane (\ref{1.7}) orthogonal to the vector ${\bf a}$, describing a circle around ${\bf a}$.

\section{Free asymmetrical body can not experience a regular precession.}\label{3RP}

Let us consider a free body that began its motion with an arbitrary initial orientation at the instant $t=0$. Its principal moments of inertia relative to the center of mass are denoted by $I_1\ne I_2\ne I_3$, while the corresponding unit vectors along the inertia axes, that form a right-handed triple, we denote ${\bf R}_1(t), {\bf R}_2(t), {\bf R}_3(t)$. We take them as the basis of the body-fixed frame. We will choose the unit vectors ${\bf e}_i$ of the Laboratory to coincide with ${\bf R}_i(t)$ at  $t=0$: ${\bf e}_i={\bf R}_i(0)$. 

Then the temporal evolution of the rotational degrees of freedom can be obtained by solving $3+9$ Euler-Poisson equations \cite{AAD_RB, AAD_23}
\begin{eqnarray}
I\dot{\boldsymbol\Omega}=[I{\boldsymbol\Omega}, {\boldsymbol\Omega}], \qquad \label{s0} \label{6.7}\\   
\dot R_{ij}=-\epsilon_{jkm}\Omega_k R_{im},   \label{s1} \label{6.8}
\end{eqnarray}
written for $3+9$ functions $R_{ij}(t)$ and $\Omega_i(t)$, considered the mutually independent variables. With our choice of the Laboratory system, the inertia tensor in these equations acquires the diagonal form: $I_{ij}=diagonal(I_1, I_2, I_3)$. The equations should be solved with universal initial data for the rotation matrix: $R_{ij}(0)=\delta_{ij}$, see (\ref{1.3}).  The initial data for $\Omega_i$ can be any three numbers: $\Omega_i(0)=\tilde\Omega_i$, they represent the initial velocity of rotation of the body.  The columns of the rotation matrix are just basis vectors of the body-fixed frame: $R(t)=({\bf R}_1(t), {\bf R}_2(t), {\bf R}_3(t))$.

{\bf Theorem 2.} Euler-Poisson equations (\ref{6.7}) and (\ref{6.8}) of an asymmetrical body do not admit a solution in the form of regular precession  (\ref{1.4}),  (\ref{1.5}) with non-zero frequencies $\alpha$ and $\beta$.  

{\bf Proof.} By construction of the angular velocity (\ref{1.1.1}), Poisson equation (\ref{6.8}) is satisfied by the functions (\ref{1.4}) and  (\ref{1.5}). So we only need to analyse the Euler equation (\ref{6.7}). The latter admits two integrals of motion. They are the rotational energy 
\begin{eqnarray}\label{s1.1}
I_1\Omega_1^2+I_2\Omega_2^2+I_3\Omega_3^2=const, \label{6.9}
\end{eqnarray}
and square of the angular momentum 
\begin{eqnarray}\label{s1.2}
I^2_1\Omega_1^2+I^2_2\Omega_2^2+I^2_3\Omega_3^2=const. \label{6.9}
\end{eqnarray}
Assuming that the regular precession solves the Euler equation, we can add the equalities (\ref{1.6}) and (\ref{1.7}) to the integrals of motion. Then the end of the vector ${\boldsymbol\Omega}(t)$ must lie at the intersection of these four surfaces of a three-dimensional space with the coordinates $\Omega_i$.   
The equations (\ref{1.6}) and (\ref{1.7}) mean that the end of the vector $\Omega_i$ lies on a plane curve (circle) that is the intersection of the sphere (\ref{1.6}) and the plane (\ref{1.7}), the latter being orthogonal to the given vector $a_i$. On the other hand, the equations (\ref{s1.1}) and (\ref{s1.2}) mean that the end of the vector $\Omega_i$ lies at the intersection of two ellipsoids. For an asymmetric body, among these curves, there are only two flat ones - these are the separatrices \cite{Landau_8}. But the solution to Euler-Poisson equations corresponding to the separatrices does not represent a regular precession, and describes the single-jump Dzhanibekov effect \cite{AAD_2023_13}. All other intersections of the two ellipsoids are not plane curves and therefore they cannot coincide with the circle specified by (\ref{1.6}) and (\ref{1.7}), or with a part of it. 

Hence, the intersection of our four surfaces can only be an isolated set of points. Then the vector $\Omega_i(t)$ does not depend on time, say $\Omega_i(t)=\tilde\Omega_i=const$. It must be a solution to the Euler equations, which acquire the following 
form: $(I_2-I_3)\tilde\Omega_2\tilde\Omega_3=(I_3-I_1)\tilde\Omega_1\tilde\Omega_3=(I_1-I_2)\tilde\Omega_1\tilde\Omega_2=0$. Their solutions are 
$(\tilde\Omega_1, 0, 0)$, $(0, \tilde\Omega_2, 0)$, and $(0, 0, \tilde\Omega_3)$. Then at $t=0$ we have $\omega_i(0)=\Omega_i(0)=\tilde\Omega_i$, that is vector of angular velocity initially is directed along one of the inertia axes. This implies pure rotation of the body around the inertia 
axis \cite{Landau_8, AAD_RB}. Thus, we have proven the impossibility of a regular precession with two non-zero frequencies.

The result obtained immediately extends to the cases of a free asymmetrical body with a fixed point and a heavy asymmetrical body with the center of mass taken as the fixed point. We recall \cite{AAD_RB} that equations of motion in these cases formally coincide with the equations (\ref{6.7}) and (\ref{6.8}).  Therefore, regular precession in these cases also turns out to be impossible. The case of an asymmetric gyroscope with a fixed point different from the center of mass will be considered in the next sections.

\section{Preliminary analysis of a regular precession of an asymmetric gyroscope.}\label{4RP}

Let us consider an asymmetric body with a fixed point $O$ different from its center of mass $G$, and immersed in a constant gravitational field. By placing the origin of the Laboratory system in the point $O$, the Euler-Poisson equations are written as follows \cite{AAD23_8, AAD23_5, AAD_RB}:  
\begin{eqnarray}
J\dot{\boldsymbol\Omega}=[J{\boldsymbol\Omega}, {\boldsymbol\Omega}]+f[R^T{\bf k}, {\bf z}(0)],  \qquad \mbox{where} \quad f=gL\mu, \label{3.1}\\   
\dot R_{ij}=-\epsilon_{jkm}\Omega_k R_{im}.   \qquad \qquad   \qquad  \qquad \qquad  \qquad ~    \label{3.2}
\end{eqnarray}
Here ${\bf k}$ is unit vector opposite to the direction of gravity force, ${\bf z}(0)$ is unit vector in the direction of center of mass at $t=0$, $g>0$ is the acceleration of gravity, $L$ is the distance to the center of mass, and $\mu$ represents the total mass of the body. 

We will start our analysis with arbitrarily chosen values of the constants involved: ${\bf a}$, $\alpha$, ${\bf b}$, $\beta$, ${\bf k}$, and ${\bf z}(0)$. Besides this, the initial orientation of the gyroscope (that is, the orientation of its inertia axes) is also arbitrary. The task is to determine all possible values of these quantities for which the rotation matrix (\ref{1.4}) satisfies the equations (\ref{3.1}) and (\ref{3.2}).

By construction of the angular velocity (\ref{1.1.1}), the Poisson equation (\ref{3.2}) is satisfied by the functions (\ref{1.4}) and  (\ref{1.5}). So we only need to analyse the Euler equation (\ref{3.1}).
Since the time dependence of angular velocity and rotation matrix is already given by Eqs. (\ref{1.4}), (\ref{1.1}), and (\ref{1.5}), substituting them into the Euler equation yields a system of algebraic equations for the possible values of the involved quantities. In this section, we will analyse these equations in the Laboratory system adapted with the vectors ${\bf k}$, ${\bf a}$, and ${\bf b}$. This will give us a number of necessary conditions for a regular precession. In the next section we will finish this analysis in the Laboratory system adapted with the inertia axes. 

\begin{figure}[t] \centering
\includegraphics[width=06cm]{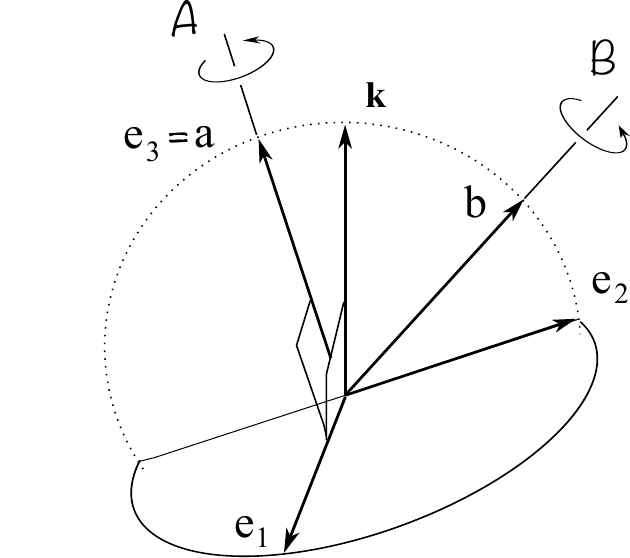}
\caption{Laboraty basis ${\bf e}_1$, ${\bf e}_2$, and ${\bf e}_3$ is adapted with ${\bf k}$, ${\bf a}$, and ${\bf b}$ at $t=0$. Except ${\bf e}_1$, 
all vectors lie on the paper sheet. Vector ${\bf k}$ is vertical.}\label{Reg_Prec_2}
\end{figure}

If the vectors ${\bf k}$ and $ {\bf b}$ are collinear, then ${\bf k}$, ${\bf b}$, and ${\bf a}$ lie in the same plane. If ${\bf k}$ and ${\bf b}$ are not collinear, then at some instant of precession the vector $R(t){\bf a}$ will be in their plane. Without loss of generality, we take this instant as the initial one, say $t=0$.  
The Laboratory system at this instant is chosen according to the following rule. Its origin we have already placed at the fixed point. Let us take  
${\bf e}_3={\bf a}$, while ${\bf e}_2$ in the plane of ${\bf k}$, ${\bf b}$, ${\bf a}$ such that $b_2>0$. Adding the vector ${\bf e}_1=[{\bf e}_2, {\bf e}_3]$ we get the right-handed Laboratory basis, see Figure \ref{Reg_Prec_2}.  
The center-of-mass vector is somehow directed in space, and therefore is not shown in the figure. 
In this Laboratory for our unit vectors we get  
\begin{eqnarray}\label{3.3}
{\bf a}=\left(\begin{array}{c} 0 \\ 0 \\ 1 \end{array}\right), \qquad   
{\bf b}=\left(\begin{array}{c} 0 \\ b_2 \\ b_3 \end{array}\right), \quad b_2>0, \quad -1<b_3<1, \qquad
{\bf k}=\left(\begin{array}{c} 0 \\ k_2 \\ k_3 \end{array}\right), \qquad 
{\bf z}(0)=\left(\begin{array}{c} z_1 \\ z_2 \\ z_3 \end{array}\right).  
\end{eqnarray}
We consider the body that began its motion with an arbitrary initial orientation at $t=0$. Since the Laboratory axes were adapted with the direction of the vectors ${\bf k}$, ${\bf a}$, and ${\bf b}$, the axes of inertia at $t=0$ are positioned arbitrarily in relation to the Laboratory. As a consequence, the inertia tensor $J_{ij}$ in the equations (\ref{3.1}) is a symmetric matrix with non-vanishing off-diagonal elements. At $t=0$, we take copies of the basis 
vectors ${\bf e}_i$, say ${\bf R}_i$, and attach them to the body. This will be our body-fixed frame. Then at future moments we 
have ${\bf R}_i(t)=R(t){\bf R}_i=R(t){\bf e}_i= (R_{1 i}(t), R_{2 i}(t), R_{3 i}(t) )^T$, that is the body-fixed frame coincides with columns of the rotation matrix.

{\bf Lemma 1.} For a regular precession, the following conditions are necessary:

1(a). Rotation and precession axes must be orthogonal: $({\bf a}, {\bf b})=0$.

1(b). Center-of-mass axis must coincide with the rotation axis: ${\EuScript Z}(t)={\EuScript A}(t)$, or ${\bf z}(t)=\pm R(t){\bf a}$.

1(c). During a regular precession, one of the inertia axes at some instant will necessarily coincide with the precession axis.

{\bf Proof.} Let us write the Euler equation as follows:  
\begin{eqnarray}\label{3.4}
{\bf E}=f{\bf T},  
\end{eqnarray}
in terms of the Euler vector  ${\bf E}\equiv J\dot{\boldsymbol\Omega}-[J{\boldsymbol\Omega}, {\boldsymbol\Omega}]$ and the 
torque ${\bf T}\equiv [R^T{\bf k}, {\bf z}(0)]$.  Using the equations (\ref{1.1}), (\ref{1.4}), and (\ref{1.5}), we find these vectors in the form of trigonometric series
\begin{eqnarray}\label{3.5}
{\bf E}=A\left(\begin{array}{c}  J_{23}\\ -J_{13} \\ 0 \end{array}\right)+
\left(\begin{array}{c} (\alpha+\beta b_3)(J_{33}-J_{22})+\alpha J_{11} \\ (2\alpha+\beta b_3)J_{12} \\ -\beta b_3 J_{13} \end{array}\right)
\beta b_2\cos\alpha t+
\left(\begin{array}{c}  -(2\alpha+\beta b_3)J_{12} \\ (\alpha+\beta b_3)(J_{11}-J_{33})-\alpha J_{22} \\ \beta b_3 J_{23} \end{array}\right)
\beta b_2\sin\alpha t + \cr
\left(\begin{array}{c}  J_{23}\\ J_{13} \\-2 J_{12} \end{array}\right) \frac{(\beta b_2)^2}{2}\cos 2\alpha t +
\left(\begin{array}{c}  J_{13}\\- J_{23} \\ 2 (J_{22}-J_{11}) \end{array}\right) \frac{(\beta b_2)^2}{2}\sin 2\alpha t; \qquad \qquad \qquad \qquad \qquad \qquad
\end{eqnarray}
\begin{eqnarray}\label{3.6}
{\bf T}=b_3 n\left(\begin{array}{c}  -z_2\\ z_1 \\ 0 \end{array}\right)+
\left(\begin{array}{c}  z_3\\ 0 \\ -z_1 \end{array}\right)b_2 n\cos\alpha t+
\left(\begin{array}{c}  0\\ -z_3 \\ z_2 \end{array}\right)b_2 n\sin\alpha t+
\left(\begin{array}{c}   z_2 \\ -z_1 \\ 0 \end{array}\right)b_2 p\cos \beta t+
\left(\begin{array}{c}   z_3 \\ 0 \\-z_1  \end{array}\right)\frac{(1+b_3)p}{2}\cos \gamma_+ t+ \cr 
\left(\begin{array}{c}   0\\ -z_3 \\ z_2  \end{array}\right)\frac{(1+b_3)p}{2}\sin \gamma_+ t+
\left(\begin{array}{c}   -z_3 \\ 0 \\ z_1  \end{array}\right)\frac{(1+b_3)p}{2}\cos \gamma_- t+
\left(\begin{array}{c}   0 \\ z_3 \\  -z_2  \end{array}\right)\frac{(1-b_3)p}{2}\cos \gamma_- t.  \qquad \qquad \qquad
\end{eqnarray}
It was denoted
\begin{eqnarray}\label{3.7}
A=\frac{(\beta b_2)^2}{2}-(\alpha+\beta b_3)^2, \quad n=({\bf b}, {\bf k})= b_2 k_2+b_3 k_3, \quad p= b_3 k_2-b_2 k_3, \quad \gamma_{+}=\alpha+\beta, \quad \gamma_{-}=\alpha-\beta.  
\end{eqnarray}

These vectors should satisfy the equation (\ref{3.4}) for any $t$; therefore, the coefficients in front of the same trigonometric functions must be equated separately. Let us recall that by construction $\alpha> 0$ and $\beta> 0$. Three cases arise that need to be considered separately. 

{\bf Case 1.} Incomparable frequencies: $\alpha\ne\beta\ne 2\alpha\ne \gamma_+\ne\gamma_-$. Then the equation (\ref{3.4}) implies, among others, the following equalities:
\begin{eqnarray}\label{3.8}
J_{12}=J_{13}=J_{23}=0, \qquad J_{22}=J_{11}. 
\end{eqnarray}
The inertia tensor in the Euler equation should be diagonal, and then $J_{22}=J_{11}$ means the equality of two among the principal moments (\ref{int.1}). The body should be dynamically symmetrical, which is not interesting to us.

{\bf Case 2.} $\beta=2\alpha$, then $\gamma_+=3\alpha$ and $\gamma_-=-\alpha$. Substituting these values of angles into the equations (\ref{3.5}) and  (\ref{3.6}), and analyzing Eq. (\ref{3.4}) with the resulting vectors, we obtain (among others) the same equalities (\ref{3.8}). The body should be symmetrical, which is not interesting to us.

{\bf Case 3.} $\beta=\alpha$, then $\gamma_+=2\alpha$ and $\gamma_-=0$. In this case, the independent equations arising from (\ref{3.4}), which do not contain the matrix element $J_{23}$, can be presented as follows:  
\begin{eqnarray}\label{3.9}
J_{12}=J_{13}=0, \qquad J_{22}=J_{11}. 
\end{eqnarray}
\begin{eqnarray}\label{3.10}
\alpha^2[J_{33}+b_3(J_{33}-J_{11})]=fnz_3, \qquad pz_1=0, \qquad pz_2=0, \qquad nz_1=0. 
\end{eqnarray}
The equations with $J_{23}$ are
\begin{eqnarray}\label{3.11}
\alpha^2[(1+b_3)(1+3b_3)-1+b_3)]J_{23}=fb_3(2nz_2-pz_3), \qquad \alpha^2b_3J_{23}=fnz_2, \qquad \alpha^2[(1-b_3)J_{23}=fpz_3. 
\end{eqnarray}
They involve the third component of the precession vector $-1<b_3<1$. If $b_3\ne 0$, the equations (\ref{3.11}) imply $J_{23}=0$. Together with (\ref{3.9}), they once again imply a symmetric body. So the only possibility left is an orthogonal precession: $b_3=0$, then ${\bf b}=(0, 1, 0)^T$. This is orthogonal to the rotation vector ${\bf a}=(0, 0, 1)^T$. Item 1(a) is proved. 

With $b_3=0$ the equations (\ref{3.11}) read as follows: $\alpha^2J_{23}=fpz_3$, $nz_2=0$. Besides this, from (\ref{3.7}) we now 
have $n=k_2$, $p=-k_3$. Taking this into account, for the orthogonal precession, the previous equations acquire the following form: 
\begin{eqnarray}\label{3.12}
J_{12}=J_{13}=0, \qquad J_{22}=J_{11}; 
\end{eqnarray}
\begin{eqnarray}\label{3.13}
\alpha^2J_{33}=fk_2z_3,  \qquad -\alpha^2J_{23}=fk_3z_3;  
\end{eqnarray}
\begin{eqnarray}\label{3.14}
k_3z_1=0, \qquad k_3z_2=0, \qquad k_2z_1=0, \qquad  k_2z_2=0. 
\end{eqnarray}

The inertia tensor is a positively defined \cite{AAD_RB}, this implies $J_{33}>0$, then (\ref{3.13}) implies $z_3\ne 0$ and $k_2\ne 0$. Then $k_3\ne 0$, otherwise (\ref{3.13}) will imply $J_{23}=0$ which, together with (\ref{3.12}) implies a symmetric body. With $k_2\ne 0$ and $k_3\ne 0$, the only possibility to satisfy the equations (\ref{3.14}) is: $z_1=z_2=0$. Then the center-of-mass vector at the initial instant is ${\bf z}(0)=(0, 0, \pm1)^T$ and therefore it is collinear with the rotation vector ${\bf a}=(0, 0, 1)^T$. Since the center-of-mass vector ${\bf z}(t)$ and the rotation vector  $R(t){\bf a}$ both are rigidly attached to the body, they will be collinear at any $t$. Item 1(b) is proved.

For the gravity vector we have obtained: ${\bf k}=(0, k_2\ne 0, k_3\ne 0)$. In particular, it can not lie on the precession axis. Therefore we arrived at a curious result noticed by Routh:  

{\bf Corollary 1.}  Center of mass of an asymmetric gyroscope cannot regularly precess about the gravity vector.

Further, let us discuss the consequences of equations (\ref{3.12}), which imply the following structure of the inertia tensor in the Laboratory basis chosen for the analysis:
\begin{eqnarray}\label{3.14.1}
J=\left(
\begin{array}{ccc}
J_{11} & 0 & 0 \\
0 &  J_{11} &  J_{23} \\
0 &  J_{23} &  J_{33} 
\end{array}
\right). 
\end{eqnarray}
To find its eigenvalues (principal moments), we need to solve the equation $\det(J-\lambda{\bf 1})=(J_{11}-\lambda)[(J_{11}-\lambda)(J_{33}-\lambda)-J_{23}^2]=0$. One solution is $\lambda=J_{11}$. The corresponding eigenvector (that is the vector along an inertia axis), obtained by solving the equation $(J-\lambda{\bf 1}){\bf s}=0$, is ${\bf s}=(s_1, 0, 0)^T$. Comparing this with ${\bf a}=( 0, 0, 1)^T$ and ${\bf b}=( 0, 1, 0, )^T$, we conclude that at $t=0$ one of the inertia axes is orthogonal to the plane of mutually orthogonal rotation and precession axes. Equivalenply, at $t=0$ it is collinear 
with ${\bf e}_1$. 

Let us consider the evolution of this axis in the process of regular precession: ${\bf s}(t)=R(t){\bf s}=R^{\bf b}(\alpha t)[R^{\bf a}(\alpha t){\bf s}]$. But the 
vector $R^{\bf a}(\alpha t){\bf s}$ is ${\bf s}$ rotated around ${\bf a}$, so at some instant, say $t_2$, it will necessarily coincide with ${\bf b}$, and 
then ${\bf s}(t_2)=R^{\bf b}(\alpha t_2){\bf b}={\bf b}$. That is, at the instant $t_2$, this inertia axis coincides with the precession axis. 
Item 1(c) and so Lemma 1 are proved.

\section{Analysis of a regular precession in the Laboratory frame adapted with inertia axes.}\label{5RP}

Although the analysis of the previous section was performed in a particular Laboratory basis, the final results have been formulated in a coordinate-free language, see  Lemma 1. However, there are still equations (\ref{3.13}) whose covariant meaning was not revealed. Besides, some additional covariant information may still be contained in the equations (\ref{3.12}). So, we now continue our analysis in the Laboratory adapted with the inertia axes, where the inertia tensor becomes diagonal. This will give additional information in terms of (coordinate-independent) principal moments of inertia (\ref{int.1}). 

\begin{figure}[t] \centering
\includegraphics[width=06cm]{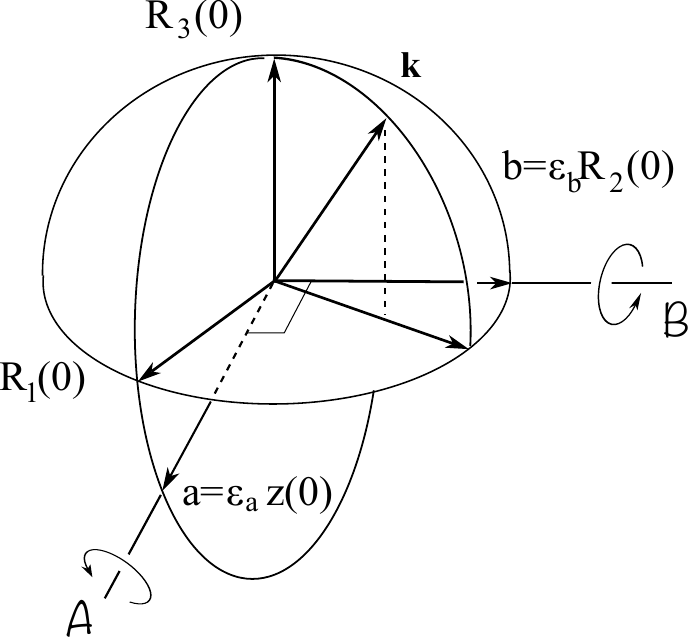}
\caption{Configuration of the gyroscope at $t=0$ taking into account Lemma 1. The vectors ${\bf R}_i(0)$ of the body-fixed frame along the inertia axes can be taken such that ${\bf R}_2(0)$ lies on the precession axis while the gravity vector lies in the first octant of the triple. The Laboratory frame is taken to coincide with the body-fixed frame at $t=0$.}\label{Reg_Prec_3}
\end{figure}

According to Item 1(c) of Lemma 1, in the process of regular precession, one of the inertia axes will pass through the precession axis. Without loss of generality, we will now take this instant as the initial one, $t=0$.  At this instant we take the right-handed triple of unit vectors ${\bf R}_i(0)$ on the inertia axes such, that ${\bf R}_2(0)$ lies on the precession axis while the gravity vector lies in the first octant of the triple. Then all its coordinates are positive numbers, $k_i>0$. Placing this basis on the paper sheet in a standard way, and taking into account the necessary conditions of Lemma 1, the configuration of our system at $t=0$ is shown in Figure \ref{Reg_Prec_3}. 
Due to the shift of the initial instant, the gravity vector now lies somewhere in the first octant, and not in any specific plane. The precession vector is collinear with ${\bf R}_2(0)$, so we write ${\bf b}=\epsilon_b {\bf R}_2(0)$, where $\epsilon_b=\pm 1$.
Similarly, ${\bf a}=\epsilon_a {\bf z}(0)$. 

The Laboratory basis we take to coincide with the body-fixed basis at $t=0$: ${\bf e}_i={\bf R}_i(0)$.  The inertia tensor in the Euler equations will now be diagonal. We enumerate the principal moments of the inertia axes in accordance with the numeration of the vectors ${\bf R}_i(t)$ fixed along them: $I_1$, $I_2$ , and $I_3$. This is some sequence of the numbers $A<B<C$ from Eq. (\ref{int.1}). Which one exactly, we still have to find out. In resume, in this Laboratory we get
\begin{eqnarray}\label{4.1}
{\bf R}_2(0)=\left(\begin{array}{c} 0 \\1 \\0 \end{array}\right), \qquad 
{\bf b}=\epsilon_b{\bf R}_2(0), \qquad 
{\bf a}=\left(\begin{array}{c} a_1\\0 \\a_3 \end{array}\right)=\epsilon_a{\bf z}(0), \qquad
{\bf k}=\left(\begin{array}{c} k_1\\ k_2 \\k_3 \end{array}\right), \quad k_i>0. 
\end{eqnarray}
In the subsequent calculations it will be convenient to represent the rotation matrix $R(t)=R^{\bf b}(\alpha t)R^{\bf a}(\alpha t)$ in terms of the vectors ${\bf a}$ and $\epsilon_b{\bf R}_2(0)$ instead of ${\bf b}$. 

We are ready to confirm the necessary and sufficient conditions described in Item I and Item II of the Introduction and then to prove Theorem 1. 

Consider the Euler equations (\ref{3.4}) in the coordinate system described above.
Using the equations (\ref{1.1}), (\ref{1.4}), (\ref{1.5}), and (\ref{4.1}) we find the Euler vector 
\begin{eqnarray}\label{4.2}
\alpha^2{\bf E}=\left(\begin{array}{c} 0 \\ -1 \\ 0 \end{array}\right)\frac12 I_{(3-1)}a_1 a_3+
\left(\begin{array}{c} a_3 \\ 0 \\ -a_1 \end{array}\right)\epsilon_b(I_1+I_{(3-2)})\cos\alpha t+
\left(\begin{array}{c} 0 \\ -1 \\ 0 \end{array}\right)\epsilon_b(I_1+I_{(3-2)})\sin\alpha t+ \cr
\left(\begin{array}{c} 0 \\ -1 \\ 0 \end{array}\right)\frac12 I_{(3-1)}a_1 a_3\cos 2\alpha t+
\left(\begin{array}{c} -I_{(3-2)}a_1 \\ 0 \\ I_{(2-1)}a_3 \end{array}\right)\frac12\sin 2\alpha t; 
\qquad \qquad \qquad \qquad 
\end{eqnarray}
and the torque
\begin{eqnarray}\label{4.3}
{\bf T}=\frac12 \left(\begin{array}{c}  \epsilon_b ({\bf a}, {\bf k})a_3 \\ k_3 a_1-k_1 a_3 \\-\epsilon_b  ({\bf a}, {\bf k})a_1 \end{array}\right)+
\left(\begin{array}{c}  a_3 k_2 \\ 0 \\-a_1 k_2 \end{array}\right)\cos \alpha t+
\left(\begin{array}{c}  0 \\ -k_2 \\ 0 \end{array}\right)\sin \alpha t+ \cr
\left(\begin{array}{c}  -\epsilon_b ({\bf a}, {\bf k})a_3 \\ k_3 a_1-k_1 a_3 \\ \epsilon_b  ({\bf a}, {\bf k})a_1 \end{array}\right)\frac12\cos 2\alpha t+
\left(\begin{array}{c}  ({\bf a}, {\bf k})a_1-k_1 \\  \epsilon_b({\bf a}, {\bf k}) \\  ({\bf a}, {\bf k})a_3-k_3 \end{array}\right)\frac12\sin 2\alpha t, 
\end{eqnarray}
where it was denoted $I_{(n-m)}\equiv I_n-I_m$.

These vectors should satisfy the equation (\ref{3.4}) for any $t$; therefore the coefficients in front of the same trigonometric functions must be equated separately.  This gives four independent relations. First, the rotation and gravity axes should be orthogonal 
\begin{eqnarray}\label{4.4.1}
({\bf a}, {\bf k})=a_1 k_1+a_3 k_3=0.
\end{eqnarray}
Second, there are three relations among the coordinates of these vectors
\begin{eqnarray}\label{4.4}
\frac{f}{\alpha^2} k_1=I_{(3-2)} a_1; \quad %\label{4.4.1} \\
\frac{f}{\alpha^2} k_2=\epsilon_b (I_{(3-2)}+I_1), \quad \mbox{this implies} \quad \epsilon_b=+1; \quad %\label{4.4.2} \\
\frac{f}{\alpha^2} k_3=-I_{(2-1)} a_3. %\label{4.4.3} 
\end{eqnarray}
The principal moments of a rigid body obey $I_3+I_1>I_2$, so $I_{(3-2)}+I_1>0$. Since $k_2>0$, the second equality from (\ref{4.4}) implies $\epsilon_b=+1$, that is the direction of precession vector ${\bf b}$ necessarily coincides with ${\bf R}_2(0)$. The rotation about ${\bf R}_2(0)$ can be only counter-clockwise. 

For the latter use we present two immediate consequences of these relations. Substituting (\ref{4.4}) into ${\bf k}^2=1$ we get 
\begin{eqnarray}\label{4.5}
\alpha^4=\frac{f^2}{I_{(3-2)} I_{(2-1)}+[I_{(3-2)}+I_1]^2}, \qquad \mbox{then} 
\quad \frac{f}{\alpha^2}=\sqrt{I_{(3-2)} I_{(2-1)}+[I_{(3-2)}+I_1]^2}.
\end{eqnarray}
The second coordinate $k_2$ represents the angle $\theta$ between the gravity and precession axes. According to (\ref{4.4}) and (\ref{4.5}) we get 
\begin{eqnarray}\label{4.6}
k_2=\cos\theta=\frac{I_{(3-2)}+I_1}{\sqrt{I_{(3-2)} I_{(2-1)}+[I_{(3-2)}+I_1]^2}}. 
\end{eqnarray}

The relations $({\bf a}, {\bf R}_2(0))=({\bf a}, {\bf k})=0$ finally fix the position of the rotation axis: it is orthogonal to the plane of the 
vectors ${\bf R}_2(0)$ and ${\bf k}$, that is, it lies along the vector $[{\bf R}_2(0), {\bf k}]$. It is convenient to introduce the unit vector 
\begin{eqnarray}\label{4.7}
{\bf n}\equiv\frac{[{\bf R}_2(0), {\bf k}]}{|[{\bf R}_2(0), {\bf k}]|}=\frac{1}{\sqrt{k_1^2+k_3^2}} \left(\begin{array}{c} k_3 \\ 0 \\ -k_1\end{array}\right), \qquad \mbox{then} \quad  {\bf a}=\pm{\bf n}.
\end{eqnarray}

{\bf Lemma 2.} It is the intermediate axis of inertia that initially coincides with the precession axis. Therefore according to (\ref{int.1}) we have  $I_2=I_B$, and  ${\bf R}_2(0)={\bf R}_B(0)$.

{\bf Proof.} By using of $k_1$ and $k_3$ from (\ref{4.4}) in (\ref{4.4.1}), and taking into account ${\bf a}^2=1$ we get 
\begin{eqnarray}\label{4.8}
I_{(3-1)}a_1^2=I_{(2-1)}, \qquad I_{(3-1)}a_1^2=I_{(2-1)}. 
\end{eqnarray}
For an asymmetric gyroscope, this implies $a_1\ne 0$ and $a_3\ne 0$. That is, the rotation vector does not lie on the axes in the plane in which it is located at $t=0$. From (\ref{4.8}) it follows that $I_{(3-1)}$, $I_{(2-1)}$, and $I_{(3-2)}$ all have the same sign. Consider these two possibilities.

{\bf (1).} If, when choosing the body-fixed basis, it turns out that they are all positive, we get 
\begin{eqnarray}\label{4.9}
I_1<I_2<I_3, \qquad \mbox{that is} \quad I_1=A, \quad I_2=B, \quad I_3=C, \quad \mbox{and} \quad
{\bf R}_1={\bf R}_A, \quad {\bf R}_2={\bf R}_B, \quad {\bf R}_3={\bf R}_C. 
\end{eqnarray}
Therefore the basis vectors ${\bf R}_1$, ${\bf R}_2$, and ${\bf R}_3$ lie on the smallest, intermediate and largest axes of inertia, 
so ${\bf R}_A, {\bf R}_B, {\bf R}_C$ is the right-handed triple.  In particular, ${\bf R}_2={\bf R}_B$. Below we will analyze the case with all $I_{(n-m)}$ negative and see that there too ${\bf R}_2={\bf R}_B$. So Lemma 2 is proved. 

Let us continue the analysis of the present case of positive $I_{(n-m)}$. From (\ref{4.4}) it follows that $a_1=a_A>0$ while $a_3=a_C<0$, then from (\ref{4.8}) the rotation vector is fixed in a unique way
\begin{eqnarray}\label{4.10}
{\bf a}=\left(\begin{array}{c} a_A \\ 0 \\ a_C \end{array}\right)=
\left(\begin{array}{c} \sqrt{\frac{B-A}{C-A}} \\ 0 \\ -\sqrt{\frac{C-B}{C-A}} \end{array}\right). 
\end{eqnarray}
In particular, its angle with smalles axis is $a_A=\cos\varphi=\sqrt{(B-A)/(C-A)}$, see Eq. (\ref{int.2}). Comparing (\ref{4.10}) with the expression (\ref{4.7}), we see that in the case (\ref{4.9}) under consideration, the rotation vector coincides with ${\bf n}$
\begin{eqnarray}\label{4.10.1}
{\bf a}=+{\bf n}=\frac{[{\bf R}_B(0), {\bf k}]}{|[{\bf R}_B(0), {\bf k}]|}. 
\end{eqnarray}

Eq. (\ref{4.6}) together with (\ref{4.9}) gives $k_2$ in terms of principal moments. The remaining two coordinates follow from (\ref{4.4}), (\ref{4.5}) and (\ref{4.10}). In the result, the position of the gravity vector is fixed in a unique way as follows:
\begin{eqnarray}\label{4.11}
{\bf k}=\left(\begin{array}{c} k_A \\ k_B \\ k_C \end{array}\right)=
\left(\begin{array}{c} \frac{C-B}{\sqrt{~}}\sqrt{\frac{B-A}{C-A}} \\ \frac{A+C-B}{\sqrt{~}} \\ \frac{B-A}{\sqrt{~}}\sqrt{\frac{C-B}{C-A}} \end{array}\right), 
\end{eqnarray}
where$\sqrt{{~}}\equiv\sqrt{(C-B)(B-A)+(A+C-B)^2}$. In particular, $k_A=\cos\theta=\frac{A+C-B}{\sqrt{~}}$, see (\ref{int.3}), 
while (\ref{4.5}) implies (\ref{int.4}).  

\begin{figure}[t] \centering
\includegraphics[width=13cm]{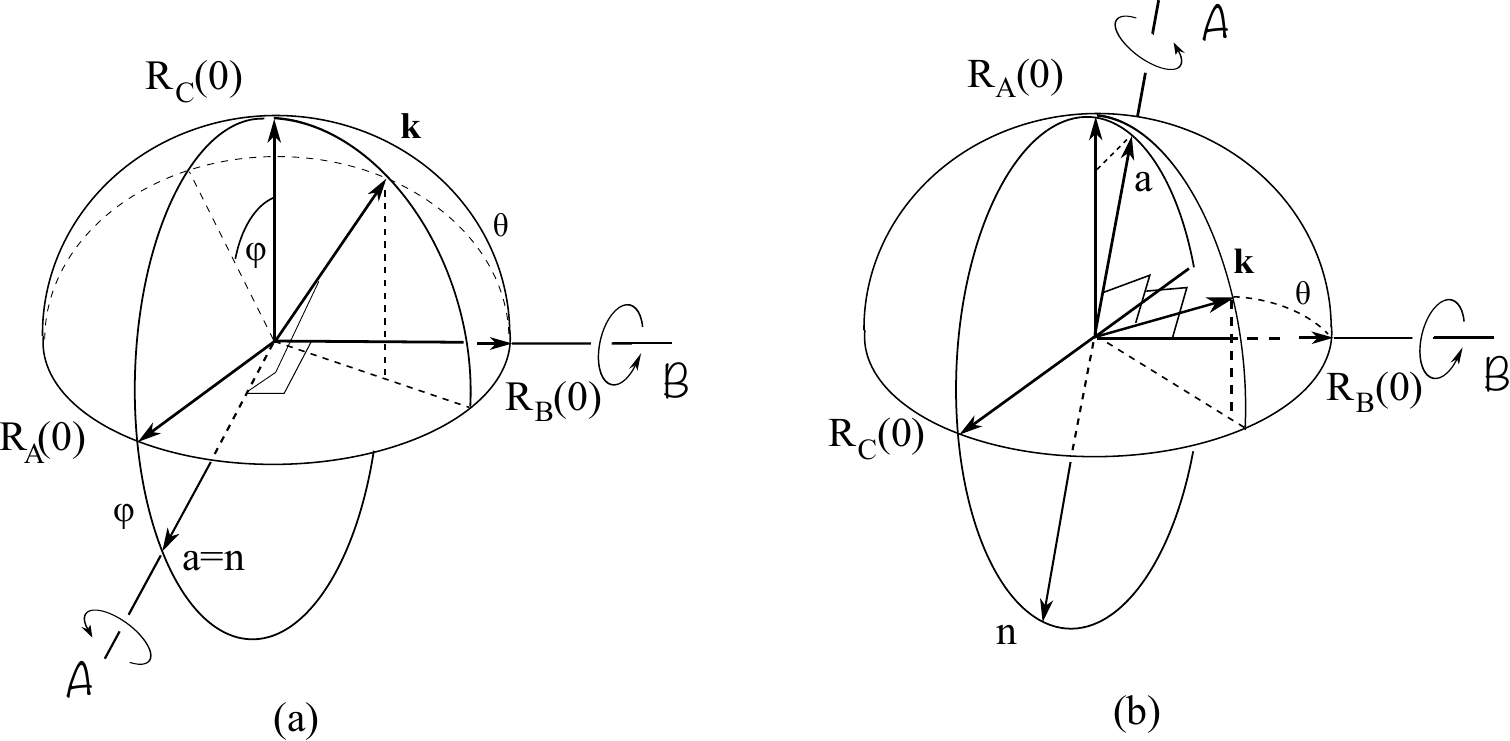}
\caption{Two possible regular precessions of an asymmetric gyroscope.}\label{Reg_Prec_4}
\end{figure}

The  Euler-Poisson equations (\ref{3.1}) and (\ref{3.2}) are now all satisfied, that is, we have found one possible regular precession of an asymmetric gyroscope, see Figure \ref{Reg_Prec_4}(a).  
This is just the counterclockwise regular precession described in Theorem 1. 

Next, let us consider the case when $I_{(n-m)}$ are all negative. 

{\bf (2).} If, when choosing the body-fixed basis, it turns out that  $I_{(3-1)}$, $I_{(2-1)}$, and $I_{(3-2)}$ are all negative, we get 
\begin{eqnarray}\label{4.12}
I_3<I_2<I_1, \qquad \mbox{that is} \quad I_3=A, \quad I_2=B, \quad I_1=C, \quad \mbox{and} \quad
{\bf R}_3={\bf R}_A, \quad {\bf R}_2={\bf R}_B, \quad {\bf R}_1={\bf R}_C. 
\end{eqnarray}
Therefore the basis vectors ${\bf R}_3$, ${\bf R}_2$, and ${\bf R}_1$ lie now on the smallest, intermediate and largest axes of inertia, and ${\bf R}_C, {\bf R}_B, {\bf R}_A$ is the right-handed triple. As before ${\bf R}_2={\bf R}_B$. 

From (\ref{4.4}) it follows that $a_1=a_C<0$ while $a_3=a_A>0$, that is, as before $a_A>0$ and $a_C<0$. Then from (\ref{4.8}) the rotation vector is fixed in a unique way
\begin{eqnarray}\label{4.13}
{\bf a}=\left(\begin{array}{c} a_C \\ 0 \\ a_A \end{array}\right)=
\left(\begin{array}{c} -\sqrt{\frac{C-B}{C-A}} \\ 0 \\\sqrt{\frac{B-A}{C-A}}  \end{array}\right). 
\end{eqnarray}
As before, $a_A=\cos\varphi=\sqrt{(B-A)/(C-A)}$, see Eq. (\ref{int.2}). Comparing (\ref{4.13}) with the expression (\ref{4.7}), we see that in the case (\ref{4.12}) under consideration, the rotation vector coincides with $-{\bf n}$
\begin{eqnarray}\label{4.14}
{\bf a}=-{\bf n}=-\frac{[{\bf R}_B(0), {\bf k}]}{|[{\bf R}_B(0), {\bf k}]|}. 
\end{eqnarray}

Eq. (\ref{4.6}) together with (\ref{4.9}) gives $k_2$ in terms of principal moments. The remaining two coordinates follow from (\ref{4.4}), (\ref{4.5}) and (\ref{4.13}). As a result, the position of the gravity vector is fixed in a unique way as follows:
\begin{eqnarray}\label{4.15}
{\bf k}=\left(\begin{array}{c} k_C, \\ k_B \\ k_A\end{array}\right)=
\left(\begin{array}{c} \frac{B-A}{\sqrt{~}}\sqrt{\frac{C-B}{C-A}} \\  \frac{A+C-B}{\sqrt{~}} \\ \frac{C-B}{\sqrt{~}}\sqrt{\frac{B-A}{C-A}} \end{array}\right), 
\end{eqnarray}
where$\sqrt{{~}}\equiv\sqrt{(C-B)(B-A)+(A+C-B)^2}$. As before, $k_A=\cos\theta=\frac{A+C-B}{\sqrt{~}}$, see (\ref{int.3}). 

The  Euler-Poisson equations (\ref{3.1}) and (\ref{3.2}) are now all satisfied, that is, we have found one more possible regular precession of an asymmetric gyroscope, see Figure \ref{Reg_Prec_4}(b).  This movement is not too different from the previous one. Indeed, let us perform a mirror reflection of the drawing \ref{Reg_Prec_4}(b) in the plane of the paper sheet, and then rotate the entire drawing $90$ degrees around  ${\bf R}_B(0)$ counter-clockwise.  The final drawing will coincide with that of Figure \ref{Reg_Prec_4}(a). But due to the reflection, the counter-clockwise rotations will turn into the clockwise ones. Therefore, the second regular precession is the first one, but the counter-clockwise rotations about ${\bf a}$ and ${\bf b}$ are replaced with the clockwise ones. This is the clockwise regular precession described in Theorem 1.

As we saw above (see Eqs. (\ref{4.10}) and (\ref{4.13}) ), for a regular precession to exist, the gyroscope's suspension point must be chosen lying on the plane of smallest and largest dynamic inertia axes so that the angle between the center-of-mass axis and the dynamic smallest axis has a strictly fixed value (\ref{int.2}). To complete the proof of Theorem 1, it remains to find out whether it is possible (and if so, how exactly) to choose the suspension point satisfying this relation. To answer this question, we need to find out how principal moments and axes of inertia behave when the suspension point shifts from the center of mass. This will be done in the next section.

\section{The behavior of principal moments and axes of inertia under shifts of the suspension point in the principal plane.}\label{6RP}

We recall that the principal moments at the center-of-mass point $G$ are denoted by $A_g<B_g<C_g$. Let us move from the center of mass to a point $O$ with the radius-vector of position $L{\bf z}$ in the direction of unit vector\footnote{From the relation (\ref{5.2}) it follows that the suspension points with the desired properties should lie in the principal plane of the axes of inertia calculated in the point $G$.}
\begin{eqnarray}\label{5.1.1}
{\bf z}=(z_A=\cos\varphi_g, ~ 0, ~ z_C=\sin\varphi_g)^T. 
\end{eqnarray}
That is, ${\bf z}$ has an angle $\varphi_g$ to the smallest axis of inertia. 
In this case, both the moments and axes at points $G$ and $O$ differ from each other. As before, the principal moments computed at $O$ are denoted by $A, B, C$. In this section, we study the dynamical principal moments and inertia axes as functions of $L$ and $\varphi_g$, and then find suspension points with the desired property (\ref{int.2}). Having chosen the Laboratory axes in the direction of inertia axes at $G$, the inertia tensor will be of diagonal form, $I_{ij}=diagonal (A_g, B_g, C_g)$. Given a point $N$ of the body, its radius-vectors ${\bf x}_N=\vec{GN}$ and ${\bf x}'_N=\vec{ON}$ are related as follows:
\begin{eqnarray}\label{5.1}
{\bf x}'_N={\bf x}_N-L{\bf z}.
\end{eqnarray}
Then the inertia tensor $I'_{ij}$ at the point $O$ through $I_{ij}$ is 
\begin{eqnarray}\label{5.2}
I'_{ij}=\sum_{N=1}^{n} m_N\left[({\bf x}'_N)^2\delta_{ij}-x'^i_Nx'^j_N\right]=I_{ij}+h\left[\delta_{ij}-z^iz^j\right], \quad 
\mbox{where} \quad  h\equiv\mu L^2. 
\end{eqnarray}
Due to the shift-induced inertia $h$, the tensor $I'$ is no longer diagonal. The matrix $\delta_{ij}-z^iz^j$ is the projector on the plane orthogonal 
to ${\bf z}$. 
Solving the equation $\det(I'-\lambda\times{\boldsymbol 1})=0$, we get eugenvalues of $I'$
\begin{eqnarray}\label{5.3}
C=\frac12\left\{A_g+C_g+h+\sqrt{{\EuScript D}}\right\}, \qquad  
B=B_g+h, \qquad 
A=\frac12\left\{A_g+C_g+h-\sqrt{{\EuScript D}}\right\}.   
\end{eqnarray}
In particular, the principal moment $B$ does not depend on the angle $\varphi_g$ and is a linear function of the induced inertia $h$. $C(h)$ and $A(h)$ are increasing functions. 
The discriminant ${\EuScript D}$ in (\ref{5.3}) can be presented in various forms, all of them turn out to be useful in the intermediate calculations
\begin{eqnarray}\label{5.4}
{\EuScript D}\equiv (A_g+C_g+h)^2-4\left[A_gC_g+h\triangle_+\right]= \qquad \qquad \qquad \qquad \qquad \qquad \qquad\cr 
h^2+2(C_g+A_g-2\triangle_+ )h+(C_g-A_g)^2=
h^2+2(C_g-A_g)(z_A^2-z_C^2)h+(C_g-A_g)^2. 
\end{eqnarray}
Here and below, we use the notation 
\begin{eqnarray}\label{5.5}
\triangle_+\equiv A_gz_A^2+C_gz_C^2, \qquad  \triangle_-\equiv A_gz_C^2+C_gz_A^2,
\end{eqnarray}
%\ref{Reg_Prec_5}  
%
%
\begin{figure}[t] \centering
\includegraphics[width=06cm]{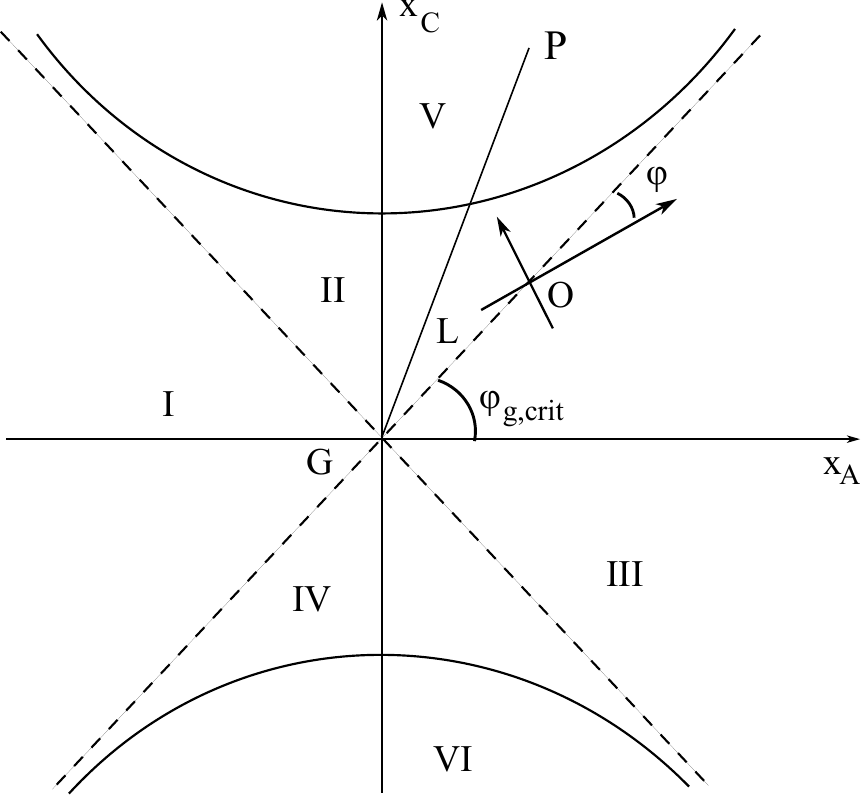}
\caption{For suspension points lying on the hyperbola, an asymmetric gyroscope becomes dynamically symmetric. For suspension points lying on the asymptotes of the hyperbola, it is possible a regular precession of an asymmetric gyroscope.}\label{Reg_Prec_5}
\end{figure}
%
%
%\ref{Reg_Prec_6}  
%
%
\begin{figure}[t] \centering
\includegraphics[width=10cm]{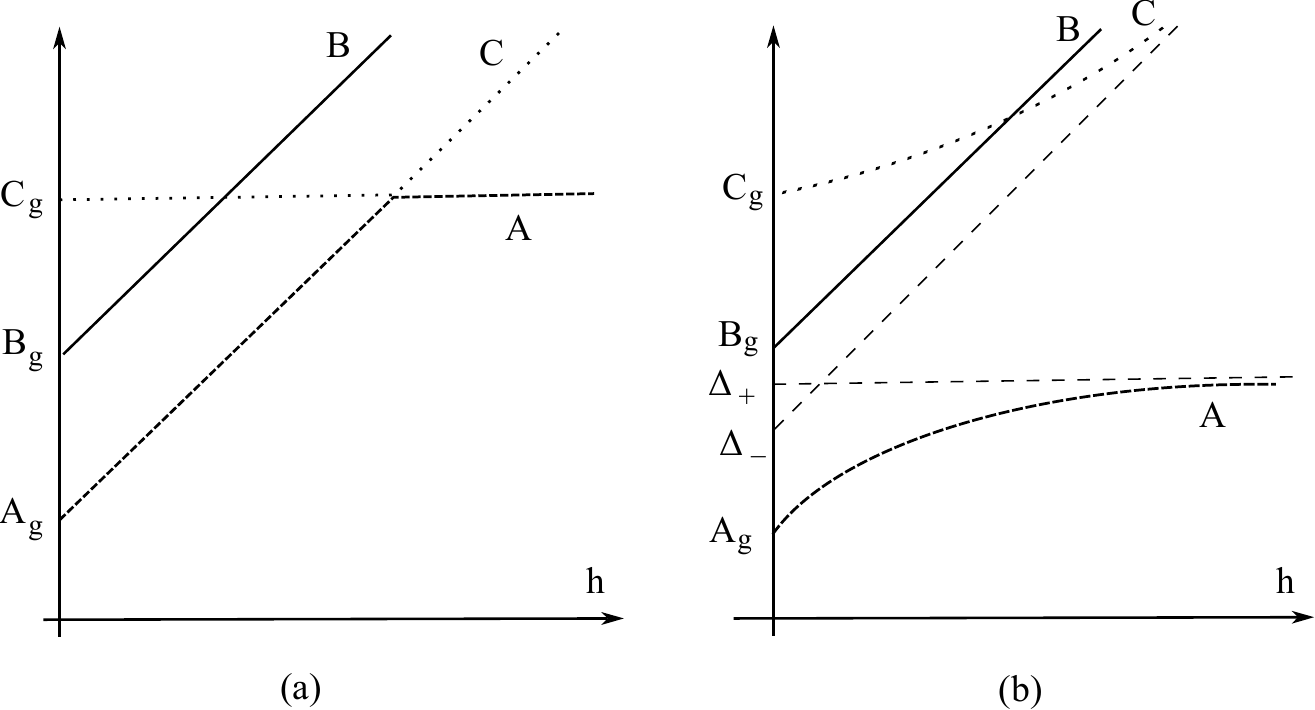}
\caption{(a) - The dependence of inertia moments on the distance $L=\sqrt{h/\mu}$  to the center of mass along a straight line passing at an 
angle $\varphi_g=\pi/2$. (b) - The dependence of inertia moments on the distance $L$ to the center of mass along a straight line passing at an 
angle $\cos^2\varphi_g<\cos^2\varphi_{g, crit}$.}\label{Reg_Prec_6}
\end{figure}
%
%
%\ref{Reg_Prec_7} 
%
\begin{figure}[t] \centering
\includegraphics[width=10cm]{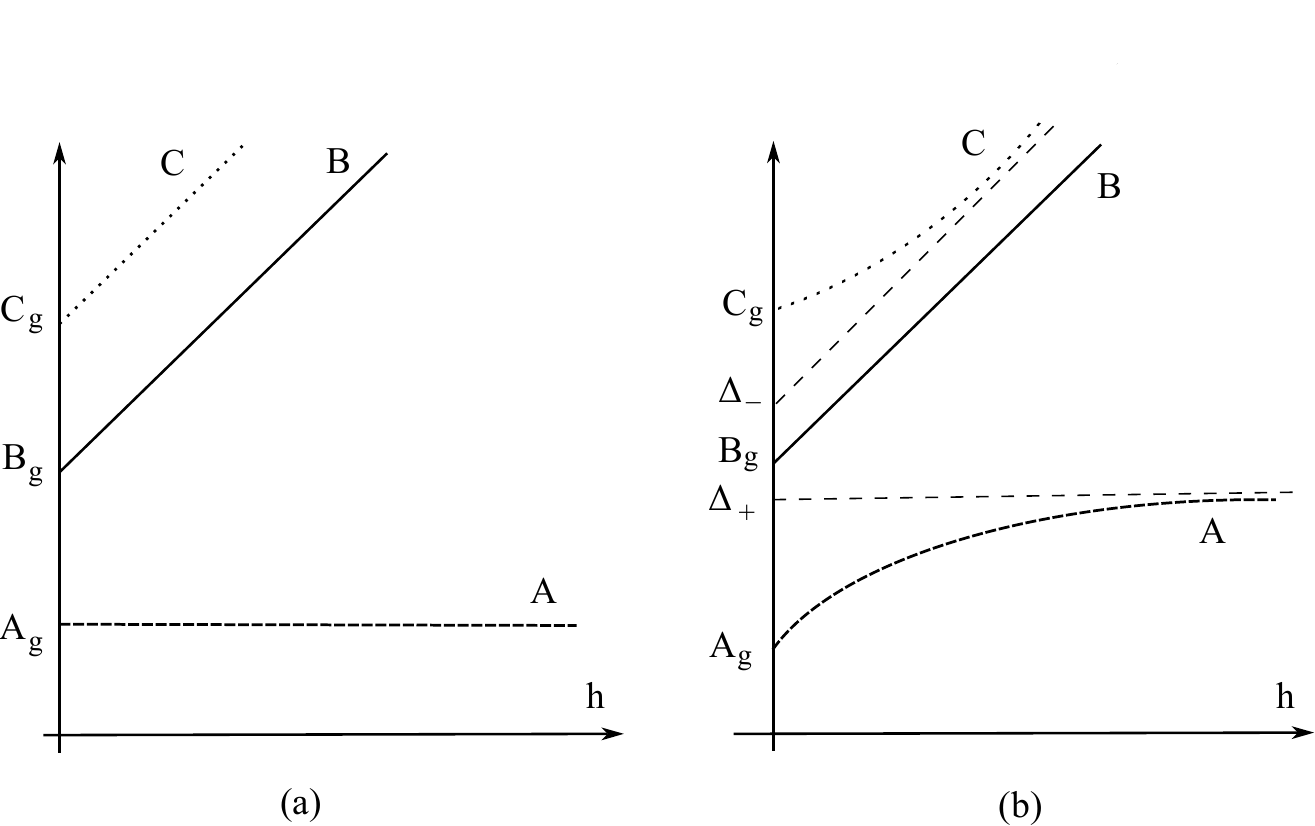}
\caption{(a) - The dependence of inertia moments on the distance $L$ to the center of mass along a straight line passing at an angle $\varphi_g=0$. (b) - The dependence of inertia moments on the distance $L$ to the center of mass along a straight line passing at an 
angle $\cos^2\varphi_g>\cos^2\varphi_{g, crit}$.}\label{Reg_Prec_7}
\end{figure}

The properties of dynamical moments as functions of a fixed point are shown in Figures \ref{Reg_Prec_5}, \ref{Reg_Prec_6}, and \ref{Reg_Prec_7}.  Figure \ref{Reg_Prec_5}  shows the plane of smallest $x_A$ and largest $x_C$ axes at the center-of-mass point $G$. Two straight lines at the critical angle 
\begin{eqnarray}\label{5.6}
z^2_A=\cos^2\varphi_{g, crit}=\frac{B_g-A_g}{C_g-A_g},
\end{eqnarray}
divide it into four regions. These straight lines are asymptotes of the hyperbola
\begin{eqnarray}\label{5.7}
-\frac{x_A^2}{(B_g-A_g)/\mu}+\frac{x_C^2}{(C_g-B_g)/\mu}=1,   
\end{eqnarray}
which further divide the area $cos^2\varphi_{g}<\cos^2\varphi_{g, crit}$ into three regions. 
By choosing any point in the regions $I$, $II$, $III$, and $IV$ as a suspension point, we get $A<B<C$. That is, the ordering of the principal moments for these points remains the same as in the center of mass.  In the regions $V$ and $VI$, we get $A<C<B$, that is, the intermediate and largest moments change places. 

Let us consider the points of a straight line $GP$ at the angle $\cos^2\varphi_{g}<\cos^2\varphi_{g, crit}$. Then at points with $0<h<h_{crit}$ we have $A<B<C$, at the point of hyperbola  
\begin{eqnarray}\label{5.8}
h_{crit}=\mu L^2=-\frac{(B_g-A_g)(C_g-B_g)}{C_g+A_g-B_g-\triangle_+},
\end{eqnarray}
the gyroscope becomes dynamically symmetrical: $A<B=C$, while at points with $h>h_{crit}$ the moments change places: $A<C<B$.

Let us consider any one of the asymptotes (\ref{5.6}). Moving the suspension point along the asymptote, the largest moment $C(h)$ asymptotically approaches the intermediate moment $B(h)$ as $h\rightarrow+\infty$.

For the suspension point lying on the hyperbola, the gyroscope is dynamically symmetrical and therefore can experience a regular precession. 

For the points lying on the asymptotes separating the regions where the interchange of the intermediate and largest moments occurs as described above, we have $A<B<C$.  As we show below, it is for these suspension points that regular precession of an asymmetric gyroscope is possible. 

When an angle $\varphi_g$ is fixed in the region $\cos^2\varphi_{g}<\cos^2\varphi_{g, crit}$, the dependence of moments on the distance $L$ to the center of mass is shown  in Figure \ref{Reg_Prec_6}. For an angle fixed in the region $\cos^2\varphi_{g}>\cos^2\varphi_{g, crit}$, the behavior of inertia moments is shown in Figure \ref{Reg_Prec_7}. 

Next, let us discuss the behavior of the dynamical axes of inertia. Let ${\bf u}$ be eigenvector of $I_{ij}$ with 
eigenvalue $B_g$, $I{\bf u}=B_g{\bf u}$. Then, according to Eq. (\ref{5.2}) we get $I'{\bf u}=(B_g+h){\bf u}$. That is, the intermediate axis does not change its orientation during the shifts (\ref{5.1.1}). Due to this, the directions of dynamical axes at the point $O$ are completely determined by the angle $\varphi$ between the smallest axis and the vector ${\bf z}$ to the center of mass (note that the smallest axis always remains smallest).  Therefore, we look for the eigenvector ${\bf v}=(v_A, ~ v_B, ~ v_C)$ corresponding to the smallest eigenvalue $A$. From the equation 
$(I'-A\times {\bf 1}){\bf v}=0$ we get $(A_g-A+hz_C^2)v_A=hz_Az_Cv_C$, and $v_B=0$, then
\begin{eqnarray}\label{5.9}
{\bf v}=\left(v_A, ~ 0, ~ \frac{A_g-A+hz_C^2}{hz_Az_C}v_A\right)^T=\left(v_A, ~ 0, ~ \frac{\sqrt{{\EuScript D}}-\alpha}{2hz_Az_C}v_A\right)^T,
\end{eqnarray}
where it was denoted 
\begin{eqnarray}\label{5.10}
\alpha\equiv-h+2hz_A^2+(C_g-A_g), \quad \mbox{then} \quad {\EuScript D}=\alpha^2+4h^2z_A^2z_C^2. 
\end{eqnarray}
The angle $\varphi$ enters into the equality $({\bf v}, ~ {\bf z})=|{\bf v}||{\bf z}|\cos\varphi$, which gives the following expression
\begin{eqnarray}\label{5.11}
2\sqrt{{\EuScript D}}\left[\sqrt{{\EuScript D}}-\alpha\right]\cos^2\varphi=z_C^2\left[2hz_A^2+\sqrt{{\EuScript D}}-\alpha\right]^2.
\end{eqnarray}
Using the equations (\ref{5.4}), (\ref{5.9}), (\ref{5.10}), and the orthogonality condition $z_A^2+z_C^2=1$, we get 
\begin{eqnarray}\label{5.12}
\cos^2\varphi=\frac12+\frac12\frac{d\sqrt{{\EuScript D}}}{dh}=
\frac12+\frac{h+(C_g-A_g)\cos 2\varphi_g}{2\sqrt{h^2+2h(C_g-A_g)\cos 2\varphi_g+(C_g-A_g)^2}}. 
\end{eqnarray}

This equality implies that when the suspension point moves away from the center of mass at an angle $\varphi_g$, the smallest dynamic axis rotates, and in the limit $L\rightarrow+\infty$ coincides with the center-of-mass axis. In particular, when shifting along the smallest axis $x_A$, the directions of dynamical axes do not change.

According to Eq. (\ref{int.2}), we are interested to find all points with the coordinates $(L, \varphi_g)$, for which 
\begin{eqnarray}\label{5.13}
\cos^2\varphi=\frac{B-A}{C-A}=\frac12+\frac{2B_g-A_g-C_g+h}{2\sqrt{{\EuScript D}}}.
\end{eqnarray}
Substituting this expression into (\ref{5.12}) we get 
\begin{eqnarray}\label{5.14}
z_A^2\equiv\cos^2\varphi_g=\frac{B_g-A_g}{C_g-A_g}, \quad \mbox{then} \quad z_C^2\equiv\sin^2\varphi_g=\frac{C_g-B_g}{C_g-A_g}.
\end{eqnarray}
Surprisingly enough, the final result does not depend on the distance from center of mass $L=\sqrt{h/\mu}$. Therefore the suspension points that allow for a regular precession lie on the straight lines 
\begin{eqnarray}\label{5.15}
x_C=\pm\sqrt{\frac{C_g-B_g}{B_g-A_g}}x_A,
\end{eqnarray}
which represent the asymptotes of the hyperbola (\ref{5.7}). As we saw above, they are frontiers of the regions where, as the distance from the center of mass increases, the interchange of the intermediate and largest moments of inertia occurs. This completes the proof of Theorem 1.

\section{Discussion. Regular precessions and quantization of spin.}\label{7RP}

One characteristic property of processes with elementary particles is the discreteness of certain quantities involved in their description: quantization of energy levels (and orbital angular momentum) in an atom, quantization of the electron's spin in a magnetic field, and so on.  The theoretical description of such properties within quantum mechanics is achieved by a radical shift of ideology compared to classical physics: equations of motion for classical continuous variables are replaced by linear equations for the wave function, involving specially chosen linear operators that allow a discrete spectrum of eigenvalues. A less radical approach is used in semiclassical models of an elementary 
particle \cite{Q_1, Q_2, Q_3, Q_4, Q_5, Q_6, Q_7, Q_8, Q_9, Q_10, Q_11, Q_12, Q_13},  where we continue to work with classical continuous variables, but satisfying modified nonlinear equations. Due to the nonlinearity, these purely classical systems can possess quantum-like behavior, reproducing several important properties of elementary particles.

A remarkable example of this kind is the classical nonlinear oscillator proposed by Rashkovskiy \cite{Rash_2011, Rash_2012}, and having many properties of the quantum mechanical harmonic oscillator. The nonlinear term in the equation of Rashkovskiy oscillator can be thought of as a friction force with the friction coefficient $k$ being a function of the energy of the system
\begin{eqnarray}\label{6.1}
m\ddot q+\omega^2 q=-k(E)\dot q, \qquad \mbox{where} \quad E\equiv\frac{m}{2}\dot q^2+\frac12\omega^2q^2. 
\end{eqnarray}
The function $k(E)$ is chosen so that for a discrete set of energy values the friction vanishes, $k(E_i)=0$, leading to quantization of the energy levels in this purely classical system. The oscillator moves along classical trajectories corresponding to these conserved energy values for an infinitely long time when external disturbances are absent. At the same time the system undergoes a spontaneous transition to a lower energy level under the action of even infinitesimal disturbances, losing the energy in the form of almost discrete portions.

Concerning the quantization of spin, consider the motion of a compass needle in the Earth's magnetic field: $\dot{\bf S}=\alpha[{\bf S}, {\bf B}]$. Due to the precession of spin ${\bf S}$ about the magnetic field, its contribution into the classical energy is $({\bf S}, {\bf B})\sim\cos\theta_0$. It can be an arbitrary number depending on the initial value of the angle $0<\theta_0<\pi$ between the magnetic field and the spin. To obtain a classical dynamic system resembling  the behavior of an electron's spin, let us consider the nonlinear equations \cite{der_QS} 
\begin{eqnarray}\label{6.2}
\frac{d{\bf S}}{dt}=-\frac{e}{mc}[{\bf B}, {\bf S}]+\beta ({\bf B}, {\bf S})[\hat{\bf S}, [\hat{\bf B}, \hat{\bf S}]], \qquad 
\hat{\bf S}={\bf S}/|{\bf S}|,  \quad \hat{\bf B}={\bf B}/|{\bf B}|. 
\end{eqnarray}
Now the evolution of ${\bf S}$ consists of two motions: precession around ${\bf B}$ caused by first term, plus circular motion on the plane of precession (that is on the plane of ${\bf B}$ and ${\bf S}(t)$) caused by second term. Due to the circular motion, a vector of spin that originally had an acute angle with ${\bf B}$ lines up in the direction of ${\bf B}$, while a vector that had an obtuse angle lines up in the opposite to ${\bf B}$ direction. These two directions are stable relative to small perturbations. In the result, the nonminimal interaction forces the spin to align up or down relative to its precession axis, leading to quantization of the energy levels in this purely classical system.

An obvious drawback of these models is that quantization does not follow from the properties of particle-field interactions established in classical physics, but requires the introduction of either exotic potentials or nonminimal interactions. In other words, as in quantum mechanics, quantization is encoded in postulates, not derived. Let us now note that in the case of regular precessions, discussed in this paper, the situation is sharply different. The frequency of regular precession was derived to be rigidly fixed (i.e., quantized: either spin up or spin down) by following the standard prescriptions of classical mechanics applied to the classical physical system: an asymmetric gyroscope.  

To deepen the analogy with the electron spin, it would be interesting to investigate regular precessions for the equations of an asymmetric body in a magnetic field \cite{AAD23_7}, and consider the question of their stability. We hope to do this in the future.

\end{document}